\def\final{1} 
\if1\final
	\documentclass[apj,twocolumn,iop]{emulateapj}
\fi
\if0\final
	\documentclass[10pt,manuscript]{aastex}
\fi

\usepackage{rcs} 
\usepackage{lineno}
\usepackage{graphicx}
\usepackage{amssymb}
\usepackage{amsmath}
\usepackage{times}
\usepackage{courier}
\usepackage{xspace}

\newcommand{\fermi}{\emph{Fermi}}
\newcommand{\lat}{\emph{Fermi}-LAT}
\newcommand{\gr}{$\gamma$-ray\xspace}
\newcommand{\grs}{$\gamma$ rays\xspace}

\newcommand{\beq}{\begin{equation}}
\newcommand{\eeq}{\end{equation}}

\renewcommand{\deg}{\ensuremath{^{\circ}}\xspace}
\newcommand{\gev}{\ensuremath{\mathrm{GeV}}\xspace}
\newcommand{\mev}{\ensuremath{\mathrm{MeV}}\xspace}

\usepackage[usenames,dvipsnames]{color}

\newcommand{\edit}[1]{\textbf{#1}}
\if1\final
	\renewcommand{\edit}[1]{#1}
\fi

\def\be{\begin{equation}}
\def\ee{\end{equation}}
\def\ea{\ et al. \,}

\begin{document}

\title{Search for gamma-ray emission from the Coma Cluster with six years of Fermi-LAT data}
\shorttitle{Coma with Pass 8}
\shortauthors{Ackermann~\ea}
\slugcomment{accepted for publication in ApJ}
\author{
M.~Ackermann\altaffilmark{1},
M.~Ajello\altaffilmark{2},
A.~Albert\altaffilmark{3},
W.~B.~Atwood\altaffilmark{4},
L.~Baldini\altaffilmark{5,3},
J.~Ballet\altaffilmark{6},
G.~Barbiellini\altaffilmark{7,8},
D.~Bastieri\altaffilmark{9,10},
K.~Bechtol\altaffilmark{11},
R.~Bellazzini\altaffilmark{12}
% ...
\\
(The Fermi-LAT Collaboration)\\
and\\
Y.~Rephaeli\altaffilmark{54,55,61}
}
\altaffiltext{1}{Deutsches Elektronen Synchrotron DESY, D-15738 Zeuthen, Germany}
\altaffiltext{2}{Department of Physics and Astronomy, Clemson University, Kinard Lab of Physics, Clemson, SC 29634-0978, USA}
\altaffiltext{3}{W. W. Hansen Experimental Physics Laboratory, Kavli Institute for Particle Astrophysics and Cosmology, Department of Physics and SLAC National Accelerator Laboratory, Stanford University, Stanford, CA 94305, USA}
\altaffiltext{4}{Santa Cruz Institute for Particle Physics, Department of Physics and Department of Astronomy and Astrophysics, University of California at Santa Cruz, Santa Cruz, CA 95064, USA}
\altaffiltext{5}{Universit\`a di Pisa and Istituto Nazionale di Fisica Nucleare, Sezione di Pisa I-56127 Pisa, Italy}
\altaffiltext{6}{Laboratoire AIM, CEA-IRFU/CNRS/Universit\'e Paris Diderot, Service d'Astrophysique, CEA Saclay, F-91191 Gif sur Yvette, France}
\altaffiltext{7}{Istituto Nazionale di Fisica Nucleare, Sezione di Trieste, I-34127 Trieste, Italy}
\altaffiltext{8}{Dipartimento di Fisica, Universit\`a di Trieste, I-34127 Trieste, Italy}
\altaffiltext{9}{Istituto Nazionale di Fisica Nucleare, Sezione di Padova, I-35131 Padova, Italy}
\altaffiltext{10}{Dipartimento di Fisica e Astronomia ``G. Galilei'', Universit\`a di Padova, I-35131 Padova, Italy}
\altaffiltext{11}{Dept.  of  Physics  and  Wisconsin  IceCube  Particle  Astrophysics  Center, University  of  Wisconsin, Madison,  WI  53706, USA}
\altaffiltext{12}{Istituto Nazionale di Fisica Nucleare, Sezione di Pisa, I-56127 Pisa, Italy}
% ...
\altaffiltext{54}{Tel Aviv University, P.O. Box 39040, Tel Aviv 6997801, Israel}
\altaffiltext{55}{Center for Astrophysics and Space Sciences, University of California, San Diego, La Jolla, CA 92093-0111, USA}
% ...
\altaffiltext{61}{email: yrephaeli@ucsd.edu}
  
\begin{abstract}
  We present results from \gr observations of the Coma cluster incorporating 6 years of \lat\ data and the newly released ``Pass 8" event-level analysis. Our analysis of the region reveals low-significance residual structures within the virial radius of the cluster that are too faint for a detailed investigation with the current data. Using a likelihood approach that is free of assumptions on the spectral shape we derive upper limits on the \gr flux that is expected from energetic particle interactions in the cluster. We also consider a benchmark spatial and spectral template motivated by models in which the observed radio halo is mostly emission by secondary electrons. In this case, the median expected and observed upper limits for the flux above $100\,\mev$ are $1.7\times10^{-9}\,\mathrm{ph\,cm^{-2}\,s^{-1}}$ and $5.2\times10^{-9}\,\mathrm{ph\,cm^{-2}\,s^{-1}}$ respectively {(the latter corresponds to residual emission at the level of $1.8\sigma$)}. These bounds are comparable to or higher than predicted levels of hadronic gamma-ray emission in {cosmic-ray models with or without reacceleration of secondary electrons}, although direct comparisons are sensitive to assumptions \edit{regarding the origin and propagation mode of cosmic rays and magnetic field properties}. The minimal expected \gr flux from radio and star-forming galaxies within the Coma cluster is roughly an order of magnitude below the median sensitivity of our analysis.

\end{abstract}


\section{Introduction}\label{sec:intro}

The radiative yields of energetic cosmic rays (CRs) traversing intracluster gas in galaxy clusters and interacting with background radiation fields span a wide spectral range from radio to high-energy \grs.  Extended regions of radio emission (referred to as halos and relics) have already been observed in many clusters \citep[e.g.,][]{Ferrari:2008aa}. Compton scattering of relativistic electrons by the Cosmic Microwave Background (CMB) radiation is the dominant process for emission above 50 keV (where thermal emission from hot intracluster gas is sufficiently weak) up to $O(100)~\mev$ \citep[e.g.,][]{Brunetti:2014aa}. Searches for this non-thermal (NT) X-ray emission have not yielded conclusive results \citep[e.g.,][]{Rephaeli:2008aa,Ajello:2009aa,Wik:2009aa}.\footnote{Recent observations of the Coma and Bullet galaxy cluster with \emph{NuSTAR} only place upper limits on the intracluster component \citep{Wik:2014aa,Gastaldello:2015aa}.} At energies higher than $O(100)~\mev$ the radiative decay of neutral pions (produced in energetic proton interactions with ambient protons, henceforth referred to as p-p) is expected to dominate, if CR interactions in the intracluster gas take place at non-negligible rates.

While there is not yet observational evidence for energetic protons in clusters, the presence of relativistic electrons is well established from radio observations. Clearly, strong radio galaxies in clusters are sources of relativistic electrons: e.g., M87 in Virgo \citep{Bolton:1949aa}, NGC~4869 in Coma \citep{Willson:1970aa}, NGC~1275  in Perseus \citep{van-den-Bergh:1961aa}; {see also \citet{Dutson:2013aa} for a comprehensive search for the brightest cluster galaxies.} It is also possible that energetic particles are (re)accelerated in the intracluster medium (ICM), by, e.g., merger and accretion shocks \citep[for a recent review, see][]{Brunetti:2014aa}. From the large extent of cluster radio halos and typical synchrotron energy loss times, it is commonly thought that the emitting electrons are secondary, namely decay products of charged pions {\citep[produced in p-p interactions; see, e.g.,][for a recent review]{Brunetti:2014aa}}. Since contemporary viable models predict that the dominant \gr emission process is $\pi^{0}$ decay \citep[e.g.,][]{Brunetti:2014aa} and since the electron lifetimes to radiative losses are generally significantly shorter than the source crossing time \citep{Petrosian:2001aa}, in our work here {we do not consider} purely leptonic models. 

An analysis of 50 clusters using four years of data from the Large Area Telescope (LAT) onboard the \fermi\ satellite \citep{Ackermann2013} resulted in upper limits on the CR-induced \gr emission \citep[see also][]{2010ApJ...717L..71A,Huber:2013aa,Griffin:2014aa,Prokhorov:2014aa}. The \gr upper limits from these analyses put stringent constraints on the energy density of energetic protons in clusters, severely constraining models for acceleration by structure formation shocks \citep[reviewed by][]{Brunetti:2014aa}. 

Due to substantial interest in exploring cluster NT emission, and the availability of a longer LAT dataset with an improved event-level analysis (see below), a dedicated analysis of the most favorable cluster candidates is warranted. The Coma cluster is a natural  choice for our search, being the nearest massive cluster with a bright radio halo. Coma is also located near the North Galactic pole where the diffuse gamma-ray intensity is at a minimum. The vast array of broadband observations, including detailed measurements of the radio halo in Coma, provide a sound basis for testing theoretical models for the gamma-ray emission {\citep[e.g.][]{Gabici:2004aa,Reimer:2004aa,Berrington:2005aa,Pinzke:2010aa,Brunetti:2012aa,Pinzke:2015aa}}. \edit{Coma has been studied using the LAT \citep[e.g.,][]{2010ApJ...717L..71A,Arlen:2012aa,Han:2012aa,Prokhorov:2014ab,Zandanel:2014aa}, its predecessor, EGRET \citep[e.g.,][]{Reimer:2003aa}, as well as Cherenkov telescopes \citep[e.g.,][]{Aharonian:2009aa,Arlen:2012aa}, but has yet to be detected in MeV-to-TeV \grs.} We report here on the deepest observation of the Coma cluster covering the \mev--\gev band to date, obtained using six years of LAT data analyzed with Pass~8 \citep{Atwood:2013aa}.

We describe the data analysis procedure in \textsection{\ref{sec:analysis}} and present our results in \textsection{\ref{sec:results}}. A comparison of the results with predicted emission levels is presented in \textsection{\ref{sec:discussion}}; we briefly summarize our conclusions in \textsection{\ref{sec:conclusion}}. In this paper we use the \emph{Planck} measurement of the Hubble constant, $H_{0}=70\,\mathrm{km\,s^{-1}\,Mpc^{-1}}$ \citep{Ade:2013aa}. At a distance of $\simeq100\,\mathrm{Mpc}$ ($z=0.023$), the virial radius of the cluster of 2.0 Mpc corresponds to a subtended angle on the sky, $\theta_{200}=1.23\deg$ \citep{Ackermann2013}.

\section{Data Selection and Analysis}\label{sec:analysis}

The LAT is a pair conversion telescope with a large field of view of $\sim2.4\,\mathrm{sr}$ sensitive to \grs from $\sim20\,\mev$ up to $>300\,\gev$ \citep{Atwood:2009aa,Ackermann:2012aa}.\footnote{Owing to the improvement with Pass~8, the effective energy range has been extended both to lower and higher energies, allowing for a more efficient reconstruction of \grs up to a few TeV with respect to previous reconstruction passes. At these energies however, the statistics remain small.}
This work uses 6 years (MET 239557414 -- 428903014) of public Pass~8 LAT data. {Pass~8 is an extensive rewrite of the core reconstruction algorithms \citep{Atwood:2013aa} applied to the data taken by the LAT that is informed by in-flight performance and characteristics of the LAT, which results in a increase in \gr acceptance by 20--40\% with respect to source events reconstructed with Pass~7REP \citep{Bregeon:2013}, depending on energy (see App.~\ref{sec:pass8} for details).}
We select photons with energies from 100~\mev to 10~\gev within a region of interest (ROI) centered on the Coma cluster at $\alpha_{2000}=194.95,\delta_{2000}=27.98$ and select a square of $15\deg\times15\deg$ as our ROI.\footnote{The coordinates of the cluster center were taken from the NASA extragalactic database (NED) \url{https://ned.ipac.caltech.edu}.} 


Limiting the data selection to zenith angles less than 90\deg allows us to effectively remove photons originating from the Earth limb.\footnote{Note that this choice is conservative to minimize the degeneracy of the extended emission from the cluster and the much brighter Limb emission.} We use {\tt{gtmktime}} to select time intervals during which the LAT was in nominal science operations mode and excluded intervals coincident with \gr bursts and solar flares. We bin our data in 16 logarithmically spaced bins in energy and use a spatial binning of 0\fdg1 per pixel.

In our analysis we include all sources listed in the latest \fermi\ catalog of point-like and extended sources \citep[3FGL,][]{Ackermann:2015} along with the standard diffuse Galactic foreground emission model recommended for point source analysis. We include a spatially isotropic model accounting for the extragalactic diffuse \gr\ background and misclassified CRs. The normalizations of both Galactic and extragalactic diffuse emission models are left free in the likelihood fit. The model components are convolved with the parametrized detector response represented by the P8R2\_SOURCE\_V6 instrument response functions (IRFs). 

We make use of recent developments of the Science Tools package that incorporate the finite energy dispersion of the LAT for the likelihood analysis by convolving the point sources in our ROI with the LAT energy dispersion.\footnote{All resources discussed, including data, analysis software, source catalog and diffuse models are available from the \fermi\ Science Support Center at \url{http://fermi.gsfc.nasa.gov/ssc/}} Note that the energy dispersion is already accounted for when creating the above-mentioned diffuse models. During the likelihood fit, we allow all sources that are separated from the cluster center by less than 6\fdg5 to have a free normalization to allow for the broad PSF at the lowest energies (the 68\% containment radius of photons at normal incidence with an energy of 100~\mev is roughly 4\deg). This choice ensures that $99.9\%$ of the predicted \gr counts (integrated over energy for an $E^{-2.3}$ spectrum, {which would roughly correspond to the predicted 
cluster emission}) is contained within the chosen radius (that is substantially larger than $\theta_{200}$ of the cluster, see \textsection{\ref{sec:modeling}} for details). 

\begin{figure*}
\begin{center}
\includegraphics[width=\textwidth]{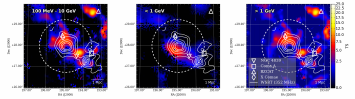}
\caption{\label{fig:tsmap}Significance map of the Coma region output from {\tt{gttsmap}} (\emph{left:} full energy range, \emph{middle:} soft band, \emph{right:} hard band). Each map has a dimension of $4\deg\times4\deg$ and a resolution of 0\fdg05 per pixel. The cyan-colored upright triangles denote positions of known blazars \citep{Massaro:2009aa} while the diamond and the square denote the position of X Comae and Coma~A, respectively (with NED-based positions). The position of NGC~4839 is marked with an inverted triangle. The dashed circle corresponds to the angle subtended by the virial radius, $\theta_{200}$. The solid contours correspond to measurements of the Coma radio halo and relic using the Westerbork Synthesis Telescope (WSRT) at 352 MHz \citep{Brown:2011aa}. The WSRT observations were convolved with a $10\arcmin$ beam and have NVSS sources removed (although the residuals of the tailed radio source in the center still contribute significantly). The lowest contour corresponds to $20~\mathrm{mJy\,beam^{-1}\,(3\sigma)}$ and increases in each contour in steps of $20~\mathrm{mJy\,beam^{-1}}$.}
\end{center}
\end{figure*}

In order to assess the completeness of our reference emission model for the Coma region (without a cluster source), we use the {\tt{gttsmap}} tool to search for any additional gamma-ray sources (Fig.~\ref{fig:tsmap}). For each pixel in the map we evaluate the test statistic ($TS$), defined as twice the likelihood ratio between the best-fit value of the alternative hypothesis, including a source and the best-fit value of the null-hypothesis (background only). In the $TS$-map a point source modeled as a power law with $\Gamma=2.3$ is tested at each pixel of the map. {We find four new point source candidates that coincide with regions with individual TS values $>25$. The details of the improved background model along with the best-fit values of all free parameters are given in App.~\ref{sec:nullmodel_fitvalues}.}

\begin{figure}
\begin{center}
\includegraphics[width=\columnwidth]{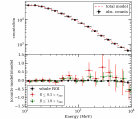}
\end{center}
\caption{\label{fig:nullfit}\emph{Top}: Observed photons in 16 energy bins along with the total number of predicted background model (comprising 3FGL sources together with extragalactic and Galactic diffuse emission templates) counts as a dashed (red) line. \emph{Bottom}: 
Spectral residuals determined from the full ROI, also evaluated within the cluster virial radius (green and red markers, respectively). For visualization purposes, the curves corresponding to the residuals within the virial radius ($R\leq0.5\times r_{200}$ and $R\leq r_{200}$) are offset by 0.1 and 0.2~dex in energy with respect to the black marker points.} 
\end{figure}

We find two residual structures whose peak $TS$ values in individual pixels (with a pixel size of $0\fdg05$) are $\sim9$ and $\sim13$, respectively. These structures, located within the virial radius, are separated from each other by about 1~Mpc at the distance of the Coma cluster, if within the cluster. We further investigate this potential excess emission by repeating the TS-map calculation in a \emph{soft}  ($E<1~\gev$) and a \emph{hard} ($E>1~\gev$) energy band. While the hard band reveals at least three distinct areas of excess emission that spatially overlap with the coordinates of X Comae, and NGC~4839, 
respectively, the soft map does not permit an immediate association with discrete sources.\footnote{Note however that we conservatively do not attempt to model the emission from these sources separately.} Moreover, when comparing our results with radio observations of the Coma cluster we find that the weak diffuse excess in the soft band roughly overlaps with that of the radio halo. In \textsection{\ref{sec:discussion}} we compare the level of this residual emission with predicted emission from likely sources in the cluster. We further investigate this apparent excess by creating spectral residuals in each energy bin $i$; for this we take the number of observed photons in each bin $c_i$ and then evaluate the number of predicted photon counts given our background-only hypothesis $m_i$ and calculate the spectral residual $r_i = (c_i - m_i)/m_{i}$. We evaluate the residuals over the full ROI as well as within 0.5 and $1.0\times \theta_{200}$ and show these results in Fig.~\ref{fig:nullfit}. While the residuals evaluated over the entire ROI are practically negligible, we note an indication of a slight excess within the cluster virial radius. 

\subsection{Cluster Spatial Modelling}\label{sec:modeling}

In this section we motivate our benchmark spatial template, hereafter referred to as the \emph{cored profile}, which is motivated by observations of the radio halo in Coma.

Evidence for NT particle populations in clusters is currently limited to measurements of extended regions of radio emission. 
The diffuse halo component in clusters (mostly intracluster emission) that remains after subtracting compact sources, has a power law spectrum with spectral flux $f_{\nu}\propto \nu^{-\alpha}$, with $\alpha \sim 1.0 - 1.5$ \citep{Ferrari:2008aa}, considerably steeper than typical Galactic spectra ($\alpha \sim 0.60 - 0.75$). The radio halo in Coma, which is centered on two dominant radio galaxies 
(NGC~4869  \& NGC~4874), 
has $\alpha \simeq 1.3$ \citep{Schlickeiser:1987aa,Thierbach:2003aa}. 
From the recent 352 MHz WSRT measurements of \citet{Brown:2011aa} the halo is $\sim 1\,\mathrm{Mpc}$ in size. The contour map in the latter paper, or, more conveniently, its conversion to an azimuthally averaged profile \citep[as plotted in Fig.~10 of][]{Planck-Collaboration:2013aa} can be parametrized by the following distribution in terms of the angular distance $\theta$ from the cluster center 
\be\label{eq:profile}
f_{r}(\theta) =\biggl[ 1 + \frac{\theta^2}{\theta_{c}^{2}}\biggr ]^{-p} \,,
\ee
with $\theta_{c} \simeq 15\arcmin$ and $p\simeq 1.4$.

The synchrotron spectral flux is proportional to the the projected (along the line of sight) value of the product $N_{e}B^{\alpha +1}$, where $N_{e}$ and $B$ are the emitting electron (column) density and locally averaged value of the magnetic field, respectively; $\alpha$ is the spectral index of the radio-synchrotron emission. Due to heavy energy losses of primary electrons (with the required energy of more than a few GeV) propagating out of their source regions, it is usually assumed  that the dominant contributions to the (observed) radio and (predicted) $\gamma$-ray emissions are from charged ($\pi^{\pm}$) and neutral ($\pi^{0}$) pion decays, respectively. The spectral intensity of the $\gamma$-ray emission is predicted to have the characteristic $\pi^{0}$ decay bump shape, unless the emission is dominated by a Compton component due to scattering of the secondary electrons off the CMB, in which case a power law with an index nearly equal to $\alpha$ would be predicted.

Whereas the spatial profile of $N_{e}$ is essentially the same as that of both $\pi^{\pm}$ and $\pi^{0}$, respective sources of the radio and $\gamma$-ray emission, the radio profile reflects also the spatial dependence of $B^{\alpha +1}$. In the hot, fully-ionized intracluster gas, electrical conductivity is high, and magnetic fields are expected to be frozen into the plasma. If so, the field spatial dependence is expected to scale as $n^{2/3}$, with $n$ being the intracluster gas density \citep{Rephaeli:1988aa}; the index in this scaling is only somewhat lower, $1/2$, if magnetic energy (rather than flux) is conserved. With a commonly used gas density profile of the form $(1 + r^{2}/r_{c}^{2})^{-3\beta/2}$, where $r_{c}$ ($\propto \theta_{c}$) is the gas core radius, and typically $\beta =2/3$, it follows that the (projected) profile of the $\gamma$-ray emission is expected to be considerably {\emph{shallower}} than the measured radio profile.\footnote{For our purposes here, this profile is sufficiently close to the more realistic density distribution deduced in the analysis of {\it Planck} Sunyaev-Zel'dovich measurements of Coma \citep{Planck-Collaboration:2013aa}.}

Taking the radio profile from Eq.~\eqref{eq:profile} and $\beta =2/3$, the profile of the $\gamma$-ray flux is predicted to be at most moderately steep, $f_{\gamma}(\theta) \propto (1 + \theta^{2}/\theta_{c}^{2})^{-1/4}$. 

In addition to the cored profile, we consider models that bracket the two extremes of possible spatial models: one that is based on the cluster being modeled as a point source (point-like) and another in which we assume a uniform distribution of predicted photons out to the virial radius (disk-like). Note that the the tabulated data for the bin-by-bin likelihoods for disk emission of varying size ($0.1\times R_{200} - 1.0\times R_{200}$) is available in a supplementary tar package.

\subsection{Likelihood Analysis}

We use an extension to the standard LAT likelihood analysis similar to the calculation of the spectral energy distribution (SED) of a source, which we refer to as the \emph{bin-by-bin likelihood}. This approach allows to calculate flux upper limits in many narrow energy bins that can then be used to easily test various broadband theoretical models without the need for a dedicated likelihood analysis \citep{Ackermann:2014aa}. Because the sensitivity of the LAT to extended emission is expected to be substantially different than for point sources, we provide a set of three SEDs corresponding to different assumed spatial templates, following the considerations from \textsection{\ref{sec:modeling}}.

For the spectral modeling, we assume the cluster emission in each energy bin to be characterized by a single power law with spectral index $\Gamma=2.3$, and calculate the profile likelihood \citep{Rolke:2005aa} of the normalization parameter. Note that the values for the nuisance parameters are determined from a global fit over the entire energy range to avoid convergence issues. We make this fit prior to constructing the bin-wise likelihood. While the choice of the index in the bin-by-bin construction is somewhat arbitrary (even though motivated by that of the measured radio spectrum of the halo), it is found to have only a marginal ($\lesssim5\%$) effect on the resulting integral flux limits (as noted in the discussion of systematics in \textsection{\ref{sec:systematics}}). 

\begin{figure*}
\begin{center}
\includegraphics[width=\textwidth]{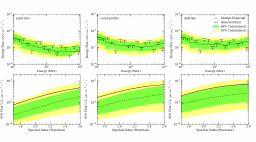}
\end{center}
\caption{\label{fig:coma_sed} \emph{Top row:} Black arrows are the observed 95\% C.L. one-sided upper limits on the integrated energy flux per energy bin when modeling the cluster as a point source \emph{left}, and as an extended source by considering either the cored profile, with a core radius of 0.3\deg (\emph{middle}), or a uniform (`disk') profile with radius $\theta_{200}=1.23\deg$ (\emph{right}). In each of these plots, the green and yellow bands indicate the 68\% and 95\% containment bands, respectively, obtained by repeating the analysis at 450 random high Galactic latitude locations in the sky. These bands provide an estimate of the overall analysis sensitivity given the assumed emission model. The dashed black curve in both panels indicates the median value. \emph{Bottom row:} Observed integral flux limits over the entire energy range from 100~\mev to 10~\gev for various spectral indices assuming a single power law in \gr\ flux for the three cluster models. The data for the full profile likelihood for each model and in each energy bin are available in the html journal.}
\end{figure*}

\section{Results}\label{sec:results}

The resultant SED for the Coma cluster is shown in Fig.~\ref{fig:coma_sed}. 
The SED was deduced by first deriving the observed 95\% confidence level one-sided upper limits, selected as the interval within which twice the difference in the log-likelihood with respect to the best-fit value of the alternative hypothesis (including the cluster) equals 2.71. These limits were then compared with those obtained by randomly selecting high Galactic latitude regions of the sky and repeating the analysis in these blank fields.\footnote{We select 450 random positions in the sky with $|b|>30\deg$ and exclude directions in which the center of the ROI coincides with either a 3FGL source or a cluster contained in the HIFLUCS catalog \citep{Reiprich:2002aa,Chen:2007}. For clusters as well as detected extended sources in 3FGL, we furthermore enforce that neither the center nor the innermost 4\deg are chosen when selecting a random sky position \citep[see, Section VI in][for a detailed discussion on the subject of blank fields]{Ackermann:2014aa}. } The observed limits are typically within the 68\% containment band, while a few bins fall in the 95\% containment band. 

The information from each bin is combined into the global log-likelihood that contains the spectral information over the whole energy range of the analysis:
\begin{equation}
 \mathcal{L}(\mu,\hat{\theta}|D)=\prod_{i}{\mathcal{L}_{i}(\mu_{i},\hat{\theta}|D_{i})}.
 \end{equation} 
In this equation we have the binned Poisson 
likelihood term $\mathcal{L}_{i}$, in each energy bin $i$, where $D_{i}$ corresponds to the observed counts in bin $i$ and $\mu_{i}$ is the normalization (the only free parameter) of the power-law with $\Gamma=2.3$ that we have assumed in each bin. $\hat{\theta}$ indicates that the nuisance parameters, $\theta$, (i.e., normalizations of free point sources and diffuse background emission components) are taken from the global fit over the entire energy range. Irrespective of the origin of the dominant emission process (be it hadronic or leptonic), in the relevant energy range explored here the spectrum can effectively be assumed to have a power-law form. Thus, we select a set of spectral indices for which we provide integral flux upper limits between 100~\mev and 10~\gev (bottom row of Fig.~\ref{fig:coma_sed}). {The maximum TS value found in our analysis is $\sim13$ for the cored profile ($\Gamma = 2.3$). When using the blank fields to assess the null-hypothesis distribution, and considering the trials factor associated with testing both for disk and point-like emission, this TS value corresponds to a global significance of $\sim1.8\sigma$. Given the low statistical significance of the residual emission, we determine integral flux upper limits.}
 
{The deduced integral flux limits in the energy range from 100~\mev to 100~\gev are $5.2\times10^{-9}\,\mathrm{ph\,cm^{-2}\,s^{-1}}$ for our benchmark model. Modeling the cluster as a point source yields a limit that is almost a factor 2 lower, $3.1\times10^{-9}\,\mathrm{ph\,cm^{-2}\,s^{-1}}$, whereas for the uniform disk model we deduce a slightly higher limit $5.8\times10^{-9}\,\mathrm{ph\,cm^{-2}\,s^{-1}}$. Note, however, that the observed limit in each case is above the blank-field median expectation, falling instead into the 68\% containment band for the disk and the 95\% containment band for point-like and cored profiles, respectively. Our main results are summarized in Table~\ref{tab:results}.}

\begin{deluxetable*}{lccc}
\tablecaption{Summary of the Likelihood Analysis Results\label{tab:results}}
\if1\final
	\tablewidth{\textwidth}
\fi
\if0\final
	\tablewidth{\textheight}
	\rotate
\fi
\tablenum{1}

\tablehead{\colhead{Spatial Model} & \colhead{Power-law index} & \colhead{$F_{\mathrm{obs}}^{95\%}(E>100~\mev)$} & \colhead{$\overline{F_{\mathrm{exp}}^{95\%}}(E>100~\mev)$} \\ 
\colhead{} & \colhead{} & \colhead{($\times10^{-9}\mathrm{ph\,cm^{-2}\,s^{-1}}$)} & \colhead{($\times10^{-9}\mathrm{ph\,cm^{-2}\,s^{-1}}$)} } 

\startdata
Cored Profile & 2.3 & 5.2 & 1.7 \\
Point Source & 2.3 & 3.1 & 1.1 \\
Disk & 2.3 & 5.8 & 2.1 \\
Cored Profile & 2.0 & 3.2 & 1.1 \\
Point Source & 2.0 & 1.7 & 0.5 \\
Disk & 2.0 & 3.8 & 1.4 \\

\enddata
\tablecomments{From left to right: spatial model assumed along with the adopted power-law index followed by the observed integral flux upper limit. The last column denotes the median expected integral flux upper limit that we determined from blank fields. All limits are calculated at 95\% C.L. Note that while we refer to $\Gamma=2.3$ as our benchmark model, we list also the values for $\Gamma=2.0$
for easier comparison with previous works (see text for details).}

\end{deluxetable*}

The limits we present here appear to be weaker than previously reported ones, most notably when comparing our results with those from the most recent analysis of the Coma cluster using \emph{Fermi}-LAT data \citep{Zandanel:2014aa}. However, the observed limits (solid black line in Fig.~\ref{fig:coma_sed}) cannot be compared naively this way because our analysis uses a larger (and different) dataset and an improved background model in terms of sources included in the ROI (see, App.~\ref{sec:pass8} for details regarding a comparison of five years of Pass 7 and Pass 8 data). Since \citet{Zandanel:2014aa} report limits with associated $TS$ values of zero, these results should instead be compared to our expected sensitivity based on the blank field study. Doing so we find that the expected sensitivity for emission from a point-like source is similar to that reported in earlier works by \citet{Zandanel:2014aa}, but the sensitivity for an extended source has improved significantly; the median expected limit for the disk-like emission for $\Gamma=2.0$ is a factor two lower than what was reported in earlier works.

\subsection{Systematic Uncertainties}\label{sec:systematics}

To assess the robustness of our results we perform a number of systematic checks and vary fiducial parameters such as the spectral or spatial binning as well as the number of free sources and repeat our analysis for these choices. In particular we investigate how these changes affect the bin-by-bin likelihood evaluations and how these changes influence our likelihood analysis for soft and hard spectra. 
We tested our results against a set of IRFs that represent minimal or maximal boundaries in the computation of the effective area and PSF within the systematic uncertainties of Pass~8 that are chosen to maximize and minimize effective area and PSF, respectively. 
To explore the uncertainties associated with the diffuse foreground emission, we follow the method in \citet{Ackermann2013} by employing a set of alternative models \citep{Ackermann:2012b,2013arXiv1304.1395D}. The total systematic uncertainties of the analysis we present here typically do not exceed 22\% for $E\gtrsim300\,\mev$. The fractional uncertainty increases up to 54\% if the full energy range ($100\,\mev \lesssim E\lesssim 10\,\gev$) is considered. For extended sources with fluxes near the current sensitivity limit of the LAT, the uncertainty from diffuse foreground modeling is at least as large as the uncertainties from the other contributors, as shown in Tab.~\ref{tab:systematics}.

\begin{deluxetable*}{lrrrr}
\if0\final
	\rotate
\fi
\tablecaption{Budget of systematic uncertainties\label{tab:systematics}}
\tablewidth{\textwidth}
\tablenum{2}
\tablehead{\colhead{Category} & \colhead{Variation} & \colhead{SED impact} & \colhead{Integral (hard)} & \colhead{Integral (soft)} \\ 
\colhead{} & \colhead{} & \colhead{} & \colhead{} & \colhead{} } 

\startdata
IRFs & bracketing MIN/MAX\tablenotemark{a} & $< 10\%\,(E\leq300\,\mev)$ & $< 3\%$ & $< 5\%$ \\
 & ($\sigma_{\mathrm{Aeff}}\sim3\%,\,\sigma_{\mathrm{PSF}}\sim5\%$) & $< 1\%\,(300\,\mev\leq E\leq 2\gev)$ &  &  \\
 &  & $< 5\%\,(E\geq2\gev)$  &  &  \\
Diffuse modeling & alt. diffuse models\tablenotemark{b}  & $< 50\%\,(E\leq300\,\mev)$ & $< 12\%$ & $< 28\%$ \\
 &  & $< 10\% (E>300\,\mev)$ &  &  \\
free radius\tablenotemark{c} & 10 deg (nominal 6.5) & $< 10 \%$ & $< 4\%$ & $< 6\%$ \\
spectral binning & $\pm50\%$ & \dots & $< 10\%$ & $< 10\%$ \\
spatial binning & $\pm50\%$ & $< 10\%$ & $< 2\%$ & $< 2\%$ \\
SED index & $\pm10\%$ & $< 2\%\,(E> 400\mev)$ & $< 5\%$ & $< 5\%$ \\
 &  & $< 10\% (E< 400\mev) $ &  &  \\
Total\tablenotemark{d} &  & $< 54\%\,(E\leq300\,\mev)$ & $< 21\%$ & $< 42\%$ \\
& & $<22\%\,(E>300\,\mev)$ & \dots & \dots \\
\enddata
\tablecomments{{For the evaluation of systematic uncertainties we consider two cases: the effect on the bin-by-bin likelihood (here the uncertainties should be applied to the upper limit reported in each bin) and the effect on the broad-band spectra assuming a hard ($\Gamma=1.6$) and soft ($\Gamma=2.6$) power-law spectrum when all spectral bins are combined (here the uncertainty is given with respect to the integral flux upper limit). }}
\tablenotetext{a}{See \textsection~5.7 and \textsection~6.5 in \citet{Ackermann:2012} for a detailed discussion of the bracketing IRF approach}
\tablenotetext{b}{See \citet{Ackermann:2012b} for the detailed model description.}
\tablenotetext{c}{Compared to the nominal choice of 6\fdg5, there are 13 more sources that are left to vary in the fit.}
\tablenotetext{d}{Values have been added in quadrature.}
\end{deluxetable*}

\section{Discussion}\label{sec:discussion}

Our analysis of 6 years of LAT Pass~8 data above 100\,\mev for the Coma cluster region does not reveal excess emission at a significance level that is large enough to claim detection of diffuse emission in the ICM. The $TS$ map in Figure~\ref{fig:tsmap} is nonetheless of appreciable interest: there seems to be residual (background-subtracted) emission from an area that overlaps partly with the Coma virial radius. It is therefore reasonable to compare the sensitivity of our analysis (as gauged through the blank field analysis) with the emission predicted from likely sources in the cluster. 

The deduced 
median integral flux upper limits for our benchmark and disk models with $\Gamma =2.3$, which corresponds to the measured index of the radio (spectral energy) flux specified above ($\alpha =1.3$), translate to the luminosity bounds of $L(>100\,\mev) \sim1.4\times 10^{42}~\mathrm{erg\,s^{-1}}$ and $L(>100\,\mev) \sim 1.8\times 10^{42}~\mathrm{erg\,s^{-1}}$, respectively, on the combined emission from all compact and extended sources in the Coma region. 

The combined 408~MHz flux density from the two central radio galaxies NGC~4869 (5C~4.81) and NGC~4874 (5C~4.85) is $2.1\times 10^{-23}~\mathrm{erg\,cm^{-2}\,s^{-1}\,Hz^{-1}}$ \citep{Willson:1970aa,Kim:1994aa}. An estimate of the combined \gr luminosity of these two galaxies can be readily made if it is assumed that the radio continuum emission is mostly due to Compton scattering of the radio-producing electrons by the CMB, a reasonable expectation in light of the conclusion reached by \citet{Abdo:2010ac} in their analysis of \lat\ measurements of Cen A, the nearby radio galaxy. If so, the total Compton luminosity, $L_{C}$, can be estimated from the measured radio (bolometric synchrotron) luminosity, $L_{S}$, by the simple scaling $L_{C} \simeq L_{S}\rho_{0}/\rho_{B}$, where $\rho_{0}$ and $ \rho_{B}=B^{2}/8\pi$ are the (present) CMB and magnetic field energy densities, respectively. With a mean field strength of $\sim 5 \,\mu$G across the the extended radio regions of NGC~4874 and NGC~4869 \citep{Feretti:1987aa, Feretti:1990aa}, we compute $L_{C} \sim 6\times10^{40}~\mathrm{erg\,s^{-1}}$, and $L(0.1-10\,\gev) \sim 2\times10^{40}~\mathrm{erg\,s^{-1}}$. Additional emission from other radio galaxies in the cluster renders this estimate a lower limit on the total \gr emission from all radio sources in Coma. 

The other discrete (`compact') sources are mostly from star-forming galaxies, whose contribution to the cluster \gr  emission was estimated by \citet{2012ApJ...755..117S} based on a scaling relation between galactic IR and \gr emission. They estimated that the superposed $0.1 - 100~\gev$ luminosity of star-forming galaxies in Coma is in the wide range of $\sim 3\times 10^{40} - 3\times 10^{42}~\mathrm{erg\,s^{-1}}$. Together, the estimates of radio and star-forming galaxies yield a lower limit on the combined emission from galaxies in Coma, $L_{\mathrm{gal}}(>100\,\mev) \sim 5\times 10^{40}\,\mathrm{erg\,s^{-1}}$. (We note in passing that the predicted \gr emission from X Comae, a background AGN at $z=0.091$ located about a degree north to the Coma center, is too weak to be of relevance for our discussion here, since  its radio flux is much weaker than the two radio galaxies within the cluster.)

Quantitative estimates of the intracluster \gr emission can only be made in the context of specific models for intracluster energetic protons and electrons producing this emission in interactions with intracluster gas (via $\pi^{0}$ decay) and Compton scattering off the CMB, respectively. A viable model must include the particle source distribution, the propagation mode, and the spatial distributions of intracluster gas and magnetic field. While models for cluster NT particles abound \citep[reviewed recently by][]{Brunetti:2014aa}, and their predicted levels of \gr emission span a wide range, we can roughly estimate a minimal level of \gr emission from the measured radio halo luminosity, avoiding the need for a detailed model (and inherent untested assumptions). The halo flux is about twice (at 408 MHz) that of the two central radio galaxies, and the mean (volume-averaged) intracluster magnetic field is at least a factor five lower than the value adopted for the central radio galaxies \citep{Bonafede:2010aa}. We would then expect that the total number of radio-emitting electrons in the halo to be considerably higher than that in the central radio galaxies. {\edit{In the context of a secondary electron origin of the measured radio halo emission,}} it can be shown that the extended $\pi^{0}$-decay \gr emission is comparable to or higher than that from the central radio galaxies. Indeed, this has recently been quantified in a detailed study by Rephaeli \& Sadeh (2016, in preparation).

The above considerations lead to the lower limit $L(>100\,\mev)\sim1\times 10^{41}\,\mathrm{erg\,s^{-1}}$ on the combined galactic and extended \gr emission from Coma. Based on this estimate the predicted {\it minimal} \gr flux is at a level that is an order of magnitude lower than our deduced upper limit. Aside from the instrument response functions, the measurement sensitivity is mainly governed by two factors: the detected (and thus modeled) discrete sources observed by \emph{Fermi}, i.e., the 3FGL, and the model for the Galactic and isotropic \gr foreground emission. \edit{Moreover, the residuals found in this work may be the first indication of cluster emission which could become more significant with further observations.}

The extrapolated sensitivity {(given by the median expectation from blank fields, scaled to a longer exposure time)} of our analysis to 10 years of LAT exposure is {comparable to recent predictions of diffusion models of CR protons with turbulent reacceleration of leptonic secondaries \citep{Brunetti:2011aa,Brunetti:2012aa}}. However, even if the exposure of the object could be doubled by the end of the \fermi\ mission, at least in the context of the analysis techniques presented here, it appears unlikely that Coma will be significantly detected in the LAT data given the current results. {This forecast, based upon the enhanced sensitivity of Pass 8 and a comparison to blank fields, is in agreement with the conclusions of \citet{Zandanel:2014aa}, who found no indications of residual emission ($TS\sim0$) in an analysis of 5 years of Pass 7 reprocessed LAT data.}
 
\section{Summary and Conclusions}\label{sec:conclusion}

The main conclusions from this paper can be summarized as follows:
\begin{enumerate}
\item{In an analysis of \grs between 100~\mev--10~\gev collected over a period of six years with the LAT and reprocessed with Pass 8, we find excess emission within the cluster virial radius; however, the statistical significance of this emission is well below the threshold to claim detection of \gr emission from the cluster. 

When spectrally analyzed, this excess emission above the background expectation can be separated into a soft and a hard component; the soft ($E<1~\gev$) appears to roughly spatially overlap 
with parts of the well measured giant radio halo, while the hard component ($E>1~\gev$) can be associated with parts of the halo.}

\item{Using a \gr template derived from the combination of Planck and WSRT observations of the radio halo, we find a maximum $TS$ value $\sim13$ for power-law emission with a spectral index $\sim2.3$ (after correcting for trial factors, this corresponds to a global significance of $~\sim1.8~\sigma$). We derive limits on the integral CR-induced \gr flux, and our observed limit excludes fluxes above $5.2\times10^{-9}~\mathrm{ph\,cm^{-2}\,s^{-1}}$. While we focus in this paper on hadronic models where the radio emission is of secondary origin, we emphasize that the results presented here are of a more universal nature; the relatively small variation of $\sim60\%$ between our benchmark model and the most extreme disk profile indicates that our results are robust with respect to the assumed spatial distribution of \grs. Our results and the bin-by-bin likelihood profiles provide the basis for comparisons with 
a variety of specific models that will help improve our understanding of CR physics in the Coma cluster.\footnote{The tabulated likelihood profiles are provided as the Data behind Figure 3.}}

\item{Based on scaling considerations and radio measurements, we derive a robust lower limit on the \gr luminosity of Coma that is a factor $\sim$ few below the median sensitivity 
given our analysis of blank fields.}
\end{enumerate}

\section*{Acknowledgments}
We thank L. Rudnick for the radio continuum maps of Coma that we used in our study. We acknowledge useful discussions with F. Zandanel and G. Brunetti.

Y. R. gratefully acknowledges the hospitality extended to him during a visit to KIPAC, where this work was initiated.    

The \textit{Fermi} LAT Collaboration acknowledges generous ongoing support from a number of agencies and institutes that have supported both the development and the operation of the LAT as well as scientific data analysis. These include the National Aeronautics and Space Administration and the Department of Energy in the United States, the Commissariat \`a l'Energie Atomique and the Centre National de la Recherche Scientifique / Institut National de Physique Nucl\'eaire et de Physique des Particules in France, the Agenzia Spaziale Italiana and the Istituto Nazionale di Fisica Nucleare in Italy, the Ministry of Education, Culture, Sports, Science and Technology (MEXT), High Energy Accelerator Research Organization (KEK) and Japan Aerospace Exploration Agency (JAXA) in Japan, and the K.~A.~Wallenberg Foundation, the Swedish Research Council and the Swedish National Space Board in Sweden.
 
Additional support for science analysis during the operations phase is gratefully acknowledged from the Istituto Nazionale di Astrofisica in Italy and the Centre National d'\'Etudes Spatiales in France.

We would like to thank the DOE SLAC National Accelerator Laboratory Computing Division for their strong support in performing the large amount of simulations and repeated analysis used in the blank field study that were necessary for this work. 

This research made use of APLpy\footnote{APLpy is an open-source plotting package for Python hosted at \url{http://aplpy.github.com}}. We acknowledge the use of the NASA/IPAC Extragalactic Database (NED) which is operated by the Jet Propulsion Laboratory, California Institute of Technology, under contract with the National Aeronautics and Space Administration and the use of HEALPix \url{http://healpix.jpl.nasa.gov/} \citep{Gorski:2005aa}.

Facility: \facility{\lat}

\bibliographystyle{apj}
\bibliography{coma}

\begin{appendix}
\section{A. Pass 8 Improvements}\label{sec:pass8}
Pass 8 is the successor of the previous event level analysis (Pass 7REP) \citep{Ackermann:2012,Bregeon:2013}. \edit{The improvements include a tree-based pattern recognition algorithm to identify and reconstruct tracks, a new energy reconstruction that uses minimum spanning trees and calorimeter clustering, which also extend the energy range at both high and low energies and an improved background rejection algorithm from the anti-coincidence shield.} Pass 8 also improved on the implementation and training of the classification trees used in the gamma-ray selection algorithms, resulting in increased gamma-ray selection efficiency with respect to Pass 7, while keeping the same background rejection power \citep{Atwood:2013aa}.\footnote{{Performance plots for Pass~8 and its comparison with Pass~7REP are provided at {\url[http://www.slac.stanford.edu/exp/glast/groups/canda/lat_Performance.htm]{http://www.slac.stanford.edu/exp/glast/groups/canda/lat\_Performance.htm}}}.} For high-level science analysis this presents a significant improvement in all metrics. Among the key improvements are a $\sim25\%$ increase in the acceptance (and more than 50\% below 100~\mev and above 300~\gev) along with an improved angular resolution. \edit{As a result the point source sensitivity in the energy range between 1~\gev and 10~\gev is enhanced by 30--40\% with respect to Pass 7REP. Similarly, the improved PSF provides increased sensitivity towards spatially extended sources.}

\begin{figure}
\begin{center}
\includegraphics[width=.5\columnwidth]{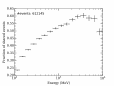}
\end{center}
\caption{\label{fig:ratio}Fraction of Pass 8 ``SOURCE" events that are present in the same integration period of five years using the roughly equivalent ``Pass 7REP" events within the Coma ROI of $15\deg\times15\deg$. Note that for Pass 7 we apply a different zenith angle cut ($100\deg$) but otherwise the basic ROI selection cuts are the same.}
\end{figure}

Due to the nature of the changes introduced with Pass~8, there is only partial overlap between the Pass 7REP and Pass 8 event samples, with the Pass 8 sample generally being the larger of the two. {The reason for this is two-fold: the changes in the event selection cause different events to pass the source selection criteria in Pass 8 than in Pass 7REP as well as selecting different residual non-photon events. The increase in \gr acceptance on the other hand increases the number of reconstructed events.} Fig.~\ref{fig:ratio} shows the fraction of shared events between our here reported Pass~8 dataset and the older Pass~7REP event sample after subjecting both datasets to the same ROI-based selection criteria. The total number of photons in our dataset is about 612k. Below $\sim3~\gev$, the fraction of shared events drops considerably, and towards the lowest energies there are about four times as many events in the Pass 8 dataset that can partially account for the differences between previously published results based on Pass 7 and the analysis we present here. 

\section{B. Details of Background Model}\label{sec:nullmodel_fitvalues}
{The basis for our background model is the 3FGL catalog of sources along with a set of diffuse templates rescaled to be used with Pass 8. We find several grid positions in the significance map (described in Section 2) that correspond to statistically significant ($TS > 25$) \gr excesses over the initial background model. The coordinates of four new point sources added to the background model are given in Table~3, along with their parameters as derived from a broad-band spectral fit. For two sources we find that a LogParabola (LP) of the form $\frac{dN}{dE}=N_{0}\left(\frac{E}{E_{b}}\right)^{-(\alpha+\beta \log (E/E_{b}))}$ gives a better fit to the data than a power law (PL); the remainder is modeled using a simple PL instead. The normalizations of each source is left free to vary in the likelihood fit.}

\begin{deluxetable}{lccccc}
\tablenum{3}
\tablecaption{Point Source Candidates in ROI\label{tab:new_ps}}
\tablehead{\colhead{Name} & \colhead{R.A.} & \colhead{Dec.} & \colhead{Model} & \colhead{Parameters} & \colhead{TS} \\ 
\colhead{} & \colhead{(\deg)} & \colhead{(\deg)} & \colhead{} & \colhead{} & \colhead{} } 
\startdata
xFGL J1904.00+3465 & 190.40 & 34.65 & PL & $\Gamma=1.88,\,E_{0}=3496\,\mev$ & 112 \\
xFGL J1927.34+3129 & 192.73 & 31.29 & LP & $\alpha=1.83,\,\beta=0.39,\,E_{b}=2407\,\mev$ & 86 \\
xFGL J2027.04+2955 & 202.70 & 29.55 & LP & $\alpha=2.57,\,\beta=0.41,\,E_{b}=1237\,\mev$ & 280 \\
xFGL J1914.49+2080 & 191.45 & 20.80 & PL & $\Gamma=3.45,\,E_{0}=1000\,\mev$ & 1220 \\
\enddata
\tablecomments{From left to right: name, right ascension, declination and assumed model along with fixed spectral parameters and $TS$ values obtained in likelihood fit to the Coma ROI. All coordinates are given in J2000 epoch.}
\end{deluxetable}

{
Table~4 shows the best-fit values for the background-only model, including our new candidate sources. The associated uncertainty refers to the 68\% parabolic error reported by MIGRAD. Note that the background model also comprises sources that are outside the selected $15\deg\times15\deg$ ROI, such as the bright source 3FGL J1224.9+2122 (4C +21.35), whose parameters are fixed to the reported values in 3FGL.}

\begin{deluxetable}{lc}
\scriptsize
\tablecaption{Best-Fit Parameters of Background Model\label{tab:nullfit}}

\tablenum{4}

\tablehead{\colhead{Source Name} & \colhead{Normalization, $N_{0}$} \\ 
\colhead{} & \colhead{($\mathrm{cm^{-2}\,s^{-1}\,\mev^{-1}}$)} } 

\startdata
3FGL J1230.3+2519 & (2.11$\pm$0.09) $\times10^{-12}$ \\
3FGL J1231.7+2847 & (0.93$\pm$0.04) $\times10^{-12}$ \\
3FGL J1254.5+2210 & (1.10$\pm$0.10) $\times10^{-13}$ \\
3FGL J1258.1+3233 & (1.40$\pm$0.10) $\times10^{-12}$ \\
3FGL J1258.4+2123 & (0.30$\pm$0.10) $\times10^{-12}$ \\
3FGL J1301.5+3333 & (0.29$\pm$0.05) $\times10^{-12}$ \\
3FGL J1303.0+2435 & (2.80$\pm$0.10) $\times10^{-12}$ \\
3FGL J1310.6+2446 & (0.35$\pm$0.09) $\times10^{-13}$ \\
3FGL J1310.6+3222 & (2.60$\pm$0.05) $\times10^{-11}$ \\
3FGL J1314.8+2349 & (0.53$\pm$0.03) $\times10^{-12}$ \\
3FGL J1321.0+2215 & (4.20$\pm$0.10) $\times10^{-12}$ \\
3FGL J1323.0+2942 & (0.60$\pm$0.03) $\times10^{-12}$ \\
3FGL J1326.1+2931 & (0.40$\pm$0.30) $\times10^{-14}$ \\
3FGL J1332.8+2723 & (0.66$\pm$0.09) $\times10^{-12}$ \\
xFGL J1914.49+2080 & (1.50$\pm$0.20) $\times10^{-13}$ \\
xFGL J1904.00+3465 & (1.40$\pm$0.40) $\times10^{-14}$ \\
xFGL J1927.34+3129 & (0.60$\pm$0.10) $\times10^{-13}$ \\
xFGL J2027.04+2955 & (2.10$\pm$0.40) $\times10^{-13}$ \\
Extragalactic Diffuse\tablenotemark{a} & (1.04$\pm$0.01) \\
Galactic Diffuse\tablenotemark{a} & (1.04$\pm$0.02)
\enddata
\tablenotetext{a}{The fitted value corresponds to the overall (unit-less) normalization of an all-sky template. The nominal value is 1.0.}

\end{deluxetable}

\end{appendix}

\end{document}